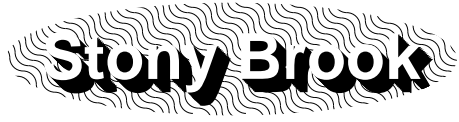

# TWO-VIERBEIN FORMALISM FOR STRING-INSPIRED AXIONIC GRAVITY

W. Siegel[1]

*Institute for Theoretical Physics*
*State University of New York, Stony Brook, NY 11794-3840*

### ABSTRACT

Using independent left and right vierbeins to describe graviton plus axion as suggested by string mechanics, O(d,d) duality can be realized linearly.

---

[1] Work supported by National Science Foundation grant PHY 9211367.

Internet address: siegel@max.physics.sunysb.edu.

# 1. INTRODUCTION

If the massless fields of the closed string are chosen to be independent of d of the D spacetime coordinates (as in dimensional reduction), the string mechanics action in this background has a global O(d,d,$\mathbf{Z}$) invariance [1-6]. This group is realized nonlinearly on the metric and axion field (two-form); in particular, the nontrivial element of the "diagonal" O(1,1,$\mathbf{Z}$) subgroup replaces metric + two-form for the d-dimensional subspace with the inverse ("$R \leftrightarrow 1/R$ duality"). This has sometimes been interpreted as illustrating the relationship between the long- and short-distance behavior of string theory; however, the fact that it is also a symmetry of the long-distance limit of closed string theory ($\alpha' \to 0$) [2] demonstrates that it is not only unrelated to short-distance behavior but also not unique to string theory. It is therefore useful to better understand this symmetry as a symmetry of massless fields to directly construct "string-inspired" theories.

The string mechanics action in background metric and axion fields can be written as
$$S = \int (\partial_+ X^m)(\partial_- X^n) e_{mn}(X), \quad e_{mn} = g_{mn} + b_{mn}$$
where $\partial_\pm$ are the usual lightlike derivatives with respect to the world-sheet coordinates (partial derivatives in the conformal gauge, or dressed up with world-sheet zweibeins more generally), which pick out the left- and right-handed modes of $X$. Normally the metric formulation of gravity can be simplified (especially in the presence of spinors) by factoring the metric as the square of the vierbein; however, the fact that $g$ and $b$ appear here only in the combination $e$ suggests that all of $e$ should be factored:
$$e_{mn} = e_{ma} e_n{}^a \quad \Rightarrow \quad S = \int (e_{ma} \partial_+ X^m)(e_n{}^a \partial_- X^n)$$
Thus $e_{ma}$ is the "left-handed" vierbein while $e_m{}^a$ is the "right-handed" one. Here $e_{ma}$ and $e_m{}^a$ are unrelated; the $a$ indices are not to be regarded as flat indices, and there is no flat-space metric $\eta_{ab}$ to raise and lower them. As a result, the local invariance on these indices is GL(D) rather than SO(D$-$1,1):
$$e_{ma} \to \Lambda_a{}^b e_{mb}, \quad e_m{}^a \to e_m{}^b \Lambda^{-1}{}_b{}^a$$
for arbitrary $\Lambda$.

This new set of fields can be used to describe any theory of gravity coupled to the axion, since the original fields can always be separated out as the symmetric and antisymmetric parts of $e_{mn}$, but the usefulness of the two-vierbein formalism is expected only for string theories, or theories with some of the properties of string theory (such as 10D supergravity). One property of string theory which survives the low-energy limit is O(d,d) duality. In this limit the restriction of the group elements to be integral can be dropped. The O(d,d) transformations are represented on $e_{mn}$ by fractional linear transformations [6]: In matrix notation,

$$e \to (\hat{\mathcal{A}}e + \hat{\mathcal{B}})(\hat{\mathcal{C}}e + \hat{\mathcal{D}})^{-1},$$

$$\hat{\mathcal{A}} = \begin{pmatrix} \mathcal{A} & 0 \\ 0 & 1 \end{pmatrix}, \quad \hat{\mathcal{B}} = \begin{pmatrix} \mathcal{B} & 0 \\ 0 & 0 \end{pmatrix}, \quad \hat{\mathcal{C}} = \begin{pmatrix} \mathcal{C} & 0 \\ 0 & 0 \end{pmatrix}, \quad \hat{\mathcal{D}} = \begin{pmatrix} \mathcal{D} & 0 \\ 0 & 1 \end{pmatrix}$$

where each vector has been divided into its d-dimensional and (D−d)-dimensional parts, respectively, and

$$O = \begin{pmatrix} \mathcal{A} & \mathcal{B} \\ \mathcal{C} & \mathcal{D} \end{pmatrix}, \quad O\eta O^T = \eta, \quad \eta = \begin{pmatrix} 0 & 1 \\ 1 & 0 \end{pmatrix}$$

is an element of O(d,d). This resembles the nonlinear transformations in O(n) nonlinear $\sigma$-models, where the fields are represented as ratios of fields that transform linearly. In this case, $e_{mn}$ is the ratio of $e_{ma}$ to $e_a{}^m$, where $e_a{}^m$ is the inverse of $e_m{}^a$ (and $e^{am}$ the inverse of $e_{ma}$). We thus consider $e_{ma}$ and $e_a{}^m$ as the basic fields, rather than $e_{ma}$ and $e_m{}^a$. The result is that the following objects transform linearly under O(d,d), as the fundamental (vector) representation ($V \to OV$):

$$\begin{pmatrix} e_{\hat{m}a} \\ e_a{}^{\hat{m}} \end{pmatrix}, \quad \begin{pmatrix} P_{\hat{m}} \\ X'^{\hat{m}} \end{pmatrix}$$

where $\hat{m}$ is the restriction of $m$ to run over just the d trivial dimensions. (In particular, the transformation $O = \begin{pmatrix} 0 & 1 \\ 1 & 0 \end{pmatrix}$ just switches the two halves.) For the other values of $m$, all these objects are invariant under O(d,d). We have included $X' = \partial X/\partial \sigma$ and $P = \delta/\delta X$ for the discussion below of these transformations for string mechanics within the hamiltonian framework [3], where duality invariance is manifest, in contrast to the usual lagrangian framework, which has half as many string variables. (The O(d,d) transformations are canonical, preserving the commutation relations of $P$ and $X'$.)

The symmetry $b \to -b$ ($\sigma \to -\sigma$) means we should be able to work equally well with $e^T$ instead of $e$. In fact, $-e^T$ transforms under the same fractional linear transformation as $e$:

$$(-e^T)' = [\hat{\mathcal{A}}(-e^T) + \hat{\mathcal{B}}][\hat{\mathcal{C}}(-e^T) + \hat{\mathcal{D}}]^{-1}$$

follows upon using $O^T \eta O = \eta$ to relate $\mathcal{A}, \mathcal{B}, \mathcal{C}, \mathcal{D}$. The fundamental O(d,d) representations would then be

$$\begin{pmatrix} -e_{\hat{m}}{}^a \\ e^{a\hat{m}} \end{pmatrix}, \quad \begin{pmatrix} P_{\hat{m}} \\ X'^{\hat{m}} \end{pmatrix}$$

The resulting duality transformations for the vierbeins are related to the earlier choice by a field-dependent GL(D) transformation: The combination of duality and GL(D) transformations on the new choice of fundamental O(d,d) representations gives the same transformation as just duality on the old choice if we pick $\Lambda$ as

$$\Lambda_a{}^b = (e_{\hat{m}a}\mathcal{A}_{\hat{n}}{}^{\hat{m}} + e_a{}^{\hat{m}}\mathcal{B}_{\hat{n}\hat{m}})(\mathcal{D}^{\hat{n}}{}_{\hat{p}}e^{b\hat{p}} - \mathcal{C}^{\hat{n}\hat{p}}e_{\hat{p}}{}^b) = -(e_{\hat{m}a}\mathcal{C}^{\hat{n}\hat{m}} + e_a{}^{\hat{m}}\mathcal{D}^{\hat{n}}{}_{\hat{m}})(\mathcal{B}_{\hat{n}\hat{p}}e^{b\hat{p}} - \mathcal{A}_{\hat{n}}{}^{\hat{p}}e_{\hat{p}}{}^b)$$

From now on we will mostly stick with the former choice, occasionally stating results for the latter choice for comparison.

This is similar to an idea of Maharana and Schwarz [7]: They gauged just GL(d). In their interpretation, fixing the dependence on d of the coordinates to be trivial was just the usual scheme for dimensional reduction, as commonly used in supergravity [8], resulting in a nonlinear $\sigma$-model which could be simplified by introducing a local internal symmetry. (Nonlinear symmetries in dimensionally reduced supergravities have long been known, but only for special D, because of the restrictions of supersymmetry.) Thus, their simplification involved treating only the scalars differently, after the usual nonlinear field redefinitions used in dimensional reduction of supergravity. (In supergravity the full global symmetry of the scalars generally doesn't appear till after the dimensional reduction, because of replacement of forms with dual forms, such as two-forms with scalars in D−d=4.) Here we take a different viewpoint: The larger local GL(D) symmetry is a symmetry of a *gravitational* theory, and provides a way of unifying the metric with the axion, in a way similar to some unified theories proposed by Einstein [9] (who did not have a symmetry to relate $g$ and $b$). It is a symmetry of gauge fields, not just scalars: This symmetry is there even if dimensional reduction is not performed. Choosing d coordinates to be trivial is then not interpreted as dimensional reduction, since those coordinates can still be finite, but as looking at particular types of solutions of the gravitational field theory (such as cosmological). Thus, we introduce the two-vierbein formalism for the complete fields even if no restriction on the coordinates is imposed, rather than for just the scalars after dimensional reduction. This additional local invariance of the two-vierbein formalism may simplify the nonlinear field redefinitions of dimensional reduction (by choice of a triangular GL gauge for both vierbeins), just as the use of a single vierbein avoids considering the redefinitions of the metric (since it is quadratic in the unredefined vierbein) but not of $b$.

These methods should also apply to supergravity theories resulting from the low-energy limit of superstrings. An enlarged vierbein was also suggested by Duff [4], but his vierbein was nonlinear, representing the fields of the nonlinear $\sigma$-model O(d,d)/O(d)⊗O(d), and does not generalize in an obvious way to all D coordinates and to the gauge fields. The idea of local symmetries for scalars being introduced before dimensional reduction was also used by de Wit and Nicolai for eleven-dimensional supergravity by making Lorentz invariance nonmanifest [10].

## 2. STRING HAMILTONIAN FORMALISM

In the hamiltonian approach one works directly with the Virasoro operators. Not only do they contain all the information in the string mechanics lagrangian (the $X$ equations of motion are irrelevant, since their time development is given by one of the Virasoro operators), but duality transformations are much simpler. The background field dependence of the Virasoro operators follows from the above lagrangian [3]:

$$\mathbf{L}_\pm = \tfrac{1}{2} g^{mn} \Pi_{\pm m} \Pi_{\pm n}, \quad \Pi_{\pm m} = P_m + (\pm g_{mn} - b_{mn}) X'^n$$

(Use of string "covariant derivatives" $\Pi$ with background $b$ has also been discussed in [11] for the generalization to Green-Schwarz strings.) General coordinate and axionic gauge transformations of the background fields are generated respectively by the two independent transformations

$$\Delta = \int d\sigma [\lambda^m(X) P_m + \lambda_m(X) X'^m]$$

acting on the Virasoro operators (either as commutators or exponentiated for finite unitary transformations). The Virasoro operators can be expressed in a form which is manifestly duality invariant: In matrix notation,

$$\mathbf{L}_\pm = \tfrac{1}{2} Z^T \eta (M \pm \eta) \eta Z$$

$$\eta = \begin{pmatrix} 0 & \delta^n_m \\ \delta^m_n & 0 \end{pmatrix} = \eta^T = \eta^{-1}$$

$$Z = \begin{pmatrix} P_m \\ X'^m \end{pmatrix}, \quad [Z(\sigma), Z(\tau)] = i\delta'(\tau - \sigma)\eta$$

$$M = \begin{pmatrix} g_{mn} - b_{mp} g^{pq} b_{qn} & b_{mp} g^{pn} \\ -g^{mp} b_{pn} & g^{mn} \end{pmatrix} = M^T = \eta M^{-1} \eta$$

$$\Delta = \int \Lambda^T \eta Z, \quad \Lambda = \begin{pmatrix} \lambda_m \\ \lambda^m \end{pmatrix}$$

When the external fields are independent of d coordinates (so the O(d,d) transformations act only on their indices and not their arguments), we then have invariance under

$$Z' = \mathcal{O}Z, \quad M' = \mathcal{O}M\mathcal{O}^T, \quad \mathcal{O}\eta\mathcal{O}^T = \eta$$

where $\mathcal{O}$ acts on just the d trivial coordinates: $\mathcal{O} = \begin{pmatrix} \hat{\mathcal{A}} & \hat{\mathcal{B}} \\ \hat{\mathcal{C}} & \hat{\mathcal{D}} \end{pmatrix}$ in the notation used earlier. ($M$ is usually used only for the scalar fields in the d trivial dimensions, but this larger $M$ gives a convenient method for expressing the O(d,d) transformations of all the fields without fractional linear transformations, and generalizes more easily to the heterotic case.)

Since $M$ is both symmetric and an element of O(D,D), it can be written as [4]

$$M = V\hat{\eta}V^T, \quad V\eta V^T = \eta, \quad \hat{\eta} = \begin{pmatrix} \eta^{mn} & 0 \\ 0 & \eta_{mn} \end{pmatrix}$$

in terms of another element $V$ of O(D,D). These two relations are covariant under the transformations

$$V' = \mathcal{O}VH, \quad H\hat{\eta}H^T = 1, \quad H\eta H^T = \eta$$

Thus, $H$ gives an O(D−1,1)⊗O(D−1,1) gauge invariance, making $V$ an element of the coset space O(D,D)/O(D−1,1)⊗O(D−1,1). Then the two O(D−1,1) vectors which are linear combinations of $V^T\eta Z$ (one vector for each of the two local O(D−1,1)'s) are the two "chiral" momenta of the string, whose squares ($= 2\mathbf{L}_\pm$) separately vanish. By choosing a non-orthonormal basis, and noting that $\hat{\eta} + \eta$ (appearing in $M + \eta = V(\hat{\eta} + \eta)V^T$) has D nonzero eigenvalues, $M$ can be expressed in the form [7]

$$M = E_a g^{ab} E_b^T - \eta, \quad g_{ab} = \tfrac{1}{2} E_a^T \eta E_b$$

where $g^{ab}$ is the inverse of $g_{ab}$ and $E_a$ is a set of O(D,D) vectors which can be identified with our two vierbeins:

$$E_a = \begin{pmatrix} e_{ma} \\ e_a{}^m \end{pmatrix}$$

$$\Rightarrow \quad g_{ab} = \tfrac{1}{2} e_{(a}{}^m e_{mb)} = e_a{}^m e_b{}^n g_{mn}$$

and $M$ is invariant under the local GL(D) transformations introduced earlier:

$$E'_a = \Lambda_a{}^b E_b \quad \Rightarrow \quad g'_{ab} = \Lambda_a{}^c \Lambda_b{}^d g_{cd}$$

Thus, the two-vierbein formalism also follows from solving the constraints on $M$ in terms of unconstrained objects (rather than elements of a group or coset space).

Using the original expressions for $g_{mn}$ and $b_{mn}$ in terms of the two vierbeins (or the expression of $M$ in terms of them), the Virasoro operators can be written as

$$\mathbf{L}_+ = \tfrac{1}{2}\Pi_a{}^T g^{ab}\Pi_b, \quad \mathbf{L}_- = \mathbf{L}_+ - Z^T\eta Z$$

$$\Pi_a = E_a{}^T \eta Z = e_a{}^m P_m + e_{ma} X'^m = e_a{}^m \Pi_{+m}$$

where $\Pi$ also is duality invariant. The gauge transformation laws for the fields follow from requiring that $\Pi$, and not just $\mathbf{L}_\pm$, transforms as $\delta\Pi_a \sim [\Delta, \Pi_a]$: They can be written in duality covariant form as

$$\delta E_{aM} = \Lambda^N \partial_N E_{aM} + E_a{}^N \partial_{[M}\Lambda_{N]}, \quad \partial_M = \begin{pmatrix} \partial_m \\ 0 \end{pmatrix} \;\Rightarrow\; \delta g_{ab} = \Lambda^M \partial_M g_{ab}$$

where $M, N$ are O(D,D) indices, raised and lowered with the O(D,D) metric $\eta_{MN}$, implicit in the matrix notation used earlier. These gauge transformations generate a local O(D,D) transformation with infintesimal parameter $\partial_{[M}\Lambda_{N]}$, as expected from the fact that the O(D,D) element $M$ is itself a representation of the gauge transformations. The separate vierbeins then transform as

$$\delta e_{ma} = (\lambda^n \partial_n e_{ma} + e_{na}\partial_m \lambda^n) + e_a{}^n \partial_{[m}\lambda_{n]}$$
$$\delta e_a{}^m = (\lambda^n \partial_n e_a{}^m - e_a{}^n \partial_n \lambda^m)$$

in addition to the GL(D) transformations described above.

Because of $b \to -b$ symmetry, we could also make the choice of opposite string chirality for expressing duality transformations:

$$M = \tilde{E}^a \tilde{g}_{ab} \tilde{E}^{bT} + \eta, \quad \tilde{g}^{ab} = -\tfrac{1}{2}\tilde{E}^{aT}\eta \tilde{E}^b$$

$$\tilde{E}^a = \begin{pmatrix} -e_m{}^a \\ e^{am} \end{pmatrix}$$

$$\tilde{g}^{ab} = \tfrac{1}{2} e^{(am} e_m{}^{b)} = e^{am} e^{bn} g_{mn} \neq g^{ab}$$

$$\mathbf{L}_- = \tfrac{1}{2}\tilde{\Pi}^{aT} \tilde{g}_{ab} \tilde{\Pi}^b, \quad \mathbf{L}_+ = \mathbf{L}_- + Z^T \eta Z$$

$$\tilde{\Pi}^a = \tilde{E}^{aT}\eta Z = e^{am} P_m - e_m{}^a X'^m = e^{am}\Pi_{-m}$$

The form of $\tilde{E}$ follows (up to a GL(D) transformation) from the fact that the consistency of these two forms of $M$ requires $\tilde{E}^{aT}\eta E_b = 0$.

## 3. GL(D) IN ORDINARY GRAVITY

Now that we have seen how duality (and gauge invariance) is manifest in the two-vierbein formalism for the background fields in string theory, we consider duality in field theory in general. Since we are considering a (gravitational) gauge theory, this analysis is simplified by the construction of covariant derivatives.

One way to define covariant derivatives is by slightly reinterpreting the approach of Cartan, who used the usual single vierbein but in a curved tangent space [12]. A convenient way to write his formalism (which he stated in the language of forms) in terms of covariant derivatives is to gauge GL(D) as above, while requiring that an independent tangent-space metric be covariantly constant:

$$\nabla_a = e_a + \omega_{ab}{}^c G_c{}^b, \quad e_a = e_a{}^m \partial_m$$

$$\nabla_a g_{bc} \equiv e_a g_{bc} + \omega_{a(bc)} = 0, \quad g_{mn} \equiv e_m{}^a e_m{}^b g_{ab}$$

$$[\nabla_a, \nabla_b] = T_{ab}{}^c \nabla_c + R_{abc}{}^d G_d{}^c, \quad [e_a, e_b] = c_{ab}{}^c e_c$$

$$T_{ab}{}^c \equiv c_{ab}{}^c + \omega_{[ab]}{}^c$$

where $G_a{}^b$ are the generators of the local GL(D) transformations and act on $a$-indices. (Thus, $\nabla_a V_b = e_a V_b + \omega_{ab}{}^c V_c$, etc. We freely raise and lower tangent space indices with the tangent-space metric.) The independent gravitational fields are the vierbein $e$ and the tangent-space metric $g$. The GL(D) connection $\omega$ is determined by $\nabla g = 0$ and the constraint that the torsion $T$ be a specified function of "matter" fields (or vanishing in the absence of matter):

$$\omega_{abc} = \tfrac{1}{2}(\tilde{c}_{bca} - \tilde{c}_{a[bc]}) + \tfrac{1}{2}(e_c g_{ab} - e_{(a} g_{b)c}), \quad \tilde{c}_{abc} = c_{abc} - T_{abc}$$

These covariant derivatives transform in the Yang-Mills way under general coordinate and local GL(D) transformations:

$$\nabla'_a = e^K \nabla_a e^{-K}, \quad K = \lambda^m \partial_m + \lambda_a{}^b G_b{}^a$$

$$g'_{ab} = e^K g_{ab}$$

(We could introduce a Christoffel term $\Gamma_{mn}{}^p G_p{}^n$ and determine it by the extra condition $\nabla_a e_b{}^m = 0$, but this condition is not covariant in this Yang-Mills sense, and would require a $\lambda_m{}^n G_n{}^m$ term in $K$ with $\lambda_m{}^n$ dependent on $\lambda^m$.) In the GL(D) gauge $g_{ab} = \eta_{ab}$, the tangent space gauge invariance is reduced to SO(D−1,1), and the usual vierbein-formalism covariant derivatives are obtained. (So $g_{ab}$ is like a Higgs field which spontaneously breaks GL(D)→SO(D−1,1).) On the other hand, in the GL(D) gauge $e_a{}^m = \delta_a^m$, $g_{ab}$ becomes the usual metric and $\omega$ becomes the usual Christoffel symbols, and we obtain the usual metric formalism. This new interpretation of Cartan's formalism (in terms of covariant derivatives with a tangent-space gauge group) requires the use of GL(D) as the gauge group of the covariant derivatives, since the usual SO(D−1,1) Lorentz gauge group does not allow for a tangent-space connection which is asymmetric in its indices.

The curvature, and in particular the curvature scalar $R \equiv R_{ab}{}^{ab}$, of any such covariant derivative $\nabla$ can be expressed in terms of the corresponding torsion-free covariant derivative $\overset{\circ}{\nabla}$ ($\overset{\circ}{T} \equiv 0$) by comparing $[\nabla, \nabla]$ with $[\overset{\circ}{\nabla}, \overset{\circ}{\nabla}]$:

$$\nabla_a = \overset{\circ}{\nabla}_a + \Delta_{ab}{}^c G_c{}^b \qquad (\Delta_{a(bc)} = 0)$$

$$\Rightarrow \quad T_{ab}{}^c = \Delta_{[ab]}{}^c, \quad R_{abc}{}^d = \overset{\circ}{R}_{abc}{}^d + \overset{\circ}{\nabla}_{[a}\Delta_{b]c}{}^d + \Delta_{[a|c}{}^e \Delta_{|b]e}{}^d$$

$$\Rightarrow \quad \Delta_{abc} = \tfrac{1}{2}(T_{a[bc]} - T_{bca})$$

$$\Rightarrow \quad R = \overset{\circ}{R} - 2\overset{\circ}{\nabla}^a T_{ab}{}^b - (T_{ab}{}^b)^2 + \tfrac{1}{4}(T_{abc})^2 - \tfrac{1}{2}T^{abc}T_{bca}$$

We also have the usual identities

$$\overset{\circ}{\nabla}_a J^a = \frac{1}{\sqrt{-g}}\partial_m(\sqrt{-g} J^m), \quad c_{ab}{}^b = -e\partial_m(e^{-1}e_a{}^m)$$

where $g = det\ g_{mn}$ (not $det\ g_{ab}$) and $e = det\ e_a{}^m$.

## 4. AXIONIC GRAVITY

This interpretation of Cartan's approach lends itself directly to the two-vierbein formalism: We choose $e_a{}^m$ to be Cartan's vierbein and $g_{ab}$ as the tangent-space metric, which has now become a composite field in terms of the vierbein $e_a{}^m$ and the "matter" field $e_{ma}$.

The field theory action for the low-energy limit of the closed, oriented, bosonic string can be written as [13,14]

$$S = \int d^D x \ \sqrt{-g} \ L$$

$$L = \phi^2(\overset{\circ}{R} - \tfrac{1}{12}g^{mn}g^{pq}g^{rs}H_{mpr}H_{nqs}) + 4g^{mn}(\partial_m \phi)(\partial_n \phi)$$

where we have absorbed the gravitational coupling into the metric, as can be done for any gravitational theory. (It then appears only through the metric's vacuum value, just as the second string coupling, which appears for the massive states, can be absorbed by the dilaton as its vacuum value.) We could make the usual rescaling $g_{mn} \to \phi^{-4/(D-2)} g_{mn}$ to rewrite the action in the form in which it appears in the bosonic sector of 10D supergravity,

$$L = \overset{\circ}{R} - \tfrac{1}{12}\phi^{8/(D-2)} H^2 - \tfrac{4}{D-2}(\partial \ln \phi)^2$$

but then duality transformations become more complicated (although the $\phi$ kinetic term now has the right sign for unitarity.) We now consider O(d,d) duality invariance of the (unscaled) low-energy action [2,5,7].

Besides $g_{ab}$, $e_a$ is duality invariant when operating on a field (with trivial dependence on $x^{\hat{m}}$). There is then the corresponding duality invariant $1 \cdot \overleftarrow{e}_a \equiv \partial_m e_a{}^m$. We will also find useful the duality invariants $f_{abc}$ and $f_{abcd}$:

$$f_{abc} \equiv \tfrac{1}{2} E_c^T \eta e_a E_b \quad\Rightarrow\quad f_{a(bc)} = e_a g_{bc}, \quad f_a{}^{(bc)} = -e_a g^{bc}$$

$$f_{abcd} \equiv \tfrac{1}{2}(e_a E_b{}^T)\eta(e_c E_d) \quad\Rightarrow\quad f_{[ab][cd]} = (c_{ab}{}^e f_{[cd]e} + c_{cd}{}^e f_{[ab]e}) - c_{ab}{}^e c_{cde}$$

$$e_{[a} f_{b]cd} - c_{ab}{}^e f_{ecd} = f_{[a|d|b]c}$$

This gives a useful expression for the axion field strength:

$$H_{mnp} \equiv \tfrac{1}{2}\partial_{[m} e_{np]} = \tfrac{1}{2}\partial_{[m} b_{np]} \quad\Rightarrow\quad H_{abc} = \tfrac{1}{2}c_{[abc]} - f_{[abc]}$$

Although general coordinate and 2-form gauge invariances $\lambda^m$ and $\lambda_m$ are not manifest in these duality invariant objects, GL(D) covariance can easily be made manifest. We first note that the GL transformation law for $f_{abc}$ allows it to be interpreted as a GL connection:

$$\delta f_{ab}{}^c = e_a \lambda_b{}^c + usual\ \lambda f\ terms$$

$$\Rightarrow\quad \omega_{ab}{}^c = -f_{ab}{}^c, \quad \tfrac{1}{2} E_c^T \eta \nabla_a E_b = 0$$

Then the previous duality invariants are replaced with the following GL(D) covariant object:

$$F_{abcd} \equiv \tfrac{1}{2}(\nabla_a E_b{}^T)\eta(\nabla_c E_d) = -\tfrac{1}{4}(e_a E_b{}^T)\eta(M-\eta)\eta(e_c E_d) = f_{abcd} - f_{ab}{}^e f_{cde}$$

$$\Rightarrow\quad F_{[ab][cd]} = -(c_{ab}{}^e - f_{[ab]}{}^e)(c_{cde} - f_{[cd]e})$$

The $M$ version of $F$ (similar to the string mechanics expression for $\mathbf{L}_-$) is GL covariant because $(M-\eta)\eta E = 0$ kills the noncovariant pieces of the transformation. Using the $e_{[a} f_{b]cd}$ identity, we also have

$$R_{abcd} = -F_{[a|d|b]c} \quad\Rightarrow\quad R = F^a{}_{[a}{}^b{}_{b]}$$

(The identity [7] $F_{acb}{}^c = \frac{1}{8}tr(e_aM)(e_bM^{-1})$ also is useful for special gauges considered in dimensional reduction.) Finally, we have the GL(D) covariantized version of $1 \cdot \overleftarrow{e}_a$:

$$1 \cdot \overleftarrow{\nabla}_a = \partial_m e_a{}^m - f_{ba}{}^b, \quad \nabla_a J^a = \partial_m J^m - (1 \cdot \overleftarrow{\nabla}_a) J^a$$

To express the action in terms of these duality invariant and GL(D) covariant objects, we first use the identities of the previous section to relate the curvature scalars:

$$T_{abc} = c_{abc} - f_{[ab]c} \quad \Rightarrow \quad H_{abc} = \tfrac{1}{2} T_{[abc]}, \quad F_{[ab][cd]} = -T_{ab}{}^e T_{cde}$$

$$\Rightarrow \quad \mathring{R} - \tfrac{1}{12} H^2 = F^a{}_{[a}{}^b{}_{b]} + F^{ab}{}_{[ab]} + 2\mathring{\nabla}^a T_{ab}{}^b + (T_{ab}{}^b)^2, \quad T_{ab}{}^b = -1 \cdot \overleftarrow{\nabla}_a - e_a ln\sqrt{-g}$$

Note that $T$, unlike $R_{abcd}$, is not duality invariant unless contracted with a derivative as $T_{ab}{}^c \nabla_c$ as it appears in $[\nabla, \nabla]$, since $\nabla$ is invariant only when it acts on a field. However, $T$ is invariant in the combination $F_{[ab][cd]}$. Furthermore, although this covariant derivative is not covariant with respect to general coordinate and $b$ gauge transformations, its covariance with respect to duality and GL(D) transformations will prove sufficient to give a simple expression for the action. (There is also a covariant derivative with $T_{abc} = H_{abc}$ implied by string mechanics [2], but it turns out not to be useful in discussing duality.)

Besides the fact that the last two, relatively simple terms in $\mathring{R} - \tfrac{1}{12} H^2$ are not duality invariant, the factor of $\sqrt{-g}$ in the integration measure is also noninvariant; the dilaton compensates for this noninvariance. (In the string quantum mechanics, it does the same for the functional integration measure for $x$, which is essentially the same thing, although in the field theory this occurs already classically.) For manifest duality and GL invariance, we define $\Phi = (-g)^{1/4} \phi$ to absorb the measure, since $g$ is not duality invariant. We then apply the identity

$$4(ue_a u^{-1} \Phi)^2 = 4(e_a \Phi)^2 + \partial_m(-2\Phi^2 e_a{}^m e^a ln\ u^2) + \Phi^2[(e_a ln\ u^2)^2 + \partial_m(2e_a{}^m e^a ln\ u^2)]$$

for the case $u = (-g)^{1/4}$. Then we find

$$[2\mathring{\nabla}^a T_{ab}{}^b + (T_{ab}{}^b)^2] + [(e_a ln\ u^2)^2 + \partial_m(2e_a{}^m e^a ln\ u^2)]$$

$$= (1 \cdot \overleftarrow{\nabla}_a)^2 + \partial_m[-2e_a{}^m(1 \cdot \overleftarrow{\nabla}^a)] = (1 \cdot \overleftarrow{\nabla}_a)^2 - 2(1 \cdot \overleftarrow{\nabla}^a \overleftarrow{\nabla}_a)$$

After an integration by parts, the action can finally be written in the simple form

$$S = \int d^D x \left\{ 4 \left[ \nabla \Phi + \tfrac{1}{2}(1 \cdot \overleftarrow{\nabla}) \Phi \right]^2 + \Phi^2 (F^a{}_{[a}{}^b{}_{b]} + F^{ab}{}_{[ab]}) \right\}$$

The $\frac{1}{2}(1\cdot\overleftarrow{\nabla})$ added to $\nabla$ on $\Phi$ is related to the fact $\Phi^2$ is the integration measure, and suggests that there should be a generalization of $\nabla$ which automatically treats $\Phi$ as a density of weight $\frac{1}{2}$. This action closely resembles the original one: It has a dilaton kinetic term, an $R = F^a{}_{[a}{}^b{}_{b]}$ curvature term, and a term $-\frac{1}{2}T_{abc}{}^2 = F^{ab}{}_{[ab]} = \frac{1}{4}(\nabla_{[a}E_{b]})^2$ analogous to the $H^2$ term. As in nonlinear $\sigma$-models, the action can be written in first-order form by making $\omega$ in $\nabla$ an independent field [7]. After making this substitution in the $\nabla$'s appearing in the definition of $F$, we find for the new $F$

$$F_{abcd} = (f_{abcd} - f_{ab}{}^e f_{cde}) + (\omega_{ab}{}^e + f_{ab}{}^e)(\omega_{cde} + f_{cde})$$

so the extra terms just fix $\omega = -f$. (There are some additional minor modifications if an independent $\omega$ is also introduced into the dilaton kinetic term.)

## 5. HETEROTIC STRING

In the hamiltonian formalism, the background formalism for the heterotic string [3] is very similar to that for the usual closed string, except that the left- and right-handed variables differ in number. (Here we consider just the bosonic sector. As usual, for representing duality we consider trivial dimensional reduction for the extra 16 dimensions, so that the gauge vectors are abelian, or we consider just a Cartan subgroup of a nonabelian group resulting from the usual compactification.) The relevant coset space is now O(D,D+n)/O(D−1,1)⊗O(D+n−1,1):

$$M = V\hat{\eta}V^T, \quad V\eta V^T = \eta$$

$$\eta = \begin{pmatrix} 0 & 1 & 0 \\ 1 & 0 & 0 \\ 0 & 0 & -1 \end{pmatrix}, \quad \hat{\eta} = \begin{pmatrix} \eta^{mn} & 0 & 0 \\ 0 & \eta_{mn} & 0 \\ 0 & 0 & -\delta_{\tilde{m}\tilde{n}} \end{pmatrix}, \quad Z = \begin{pmatrix} P_m \\ X'^m \\ \frac{1}{\sqrt{2}}(P^{\tilde{m}} + X'^{\tilde{m}}) \end{pmatrix}$$

where the new entry for $Z$ represents the chiral bosons. Now $V^T \eta Z$ for the string consists of an O(D-1,1) vector and an O(D+n-1,1) vector (D=10, D+n=26) for the left- and right-handed string momenta. By choosing a (non-orthonormal) basis, the solution to these conditions can again be expressed as

$$M = E_a g^{ab} E_b{}^T - \eta, \quad g_{ab} = \tfrac{1}{2} E_a{}^T \eta E_b, \quad E_a = \begin{pmatrix} e_{ma} \\ e_a{}^m \\ e_a{}^{\tilde{m}} \end{pmatrix}$$

(There is also an opposite chirality solution $\tilde{E}^A$, where $A$ is now a GL(D+n) index. These two choices again correspond to the number of nonvanishing eigenvalues of $\hat{\eta} \pm \eta$, as seen by performing the transformation $V^{-1}$ on $M$. For simplicity we will stick to just the GL(D) chirality given above, but similar expressions exist for the GL(D+n) chirality. The resulting local GL(D) or GL(D+n) invariance then leaves the appropriate D(D+n) components in both cases.)

The definitions of the usual gauge fields follow from their gauge transformation laws:

$$\delta E_{aM} = \Lambda^N \partial_N E_{aM} + E_a{}^N \partial_{[M} \Lambda_{N]}, \quad \partial_M = \begin{pmatrix} \partial_m \\ 0 \\ 0 \end{pmatrix}, \quad \Lambda_M = \begin{pmatrix} \lambda_m \\ \lambda^m \\ \lambda^{\tilde{m}} \end{pmatrix}$$

$$\Rightarrow \quad \delta e_{ma} = (\lambda^n \partial_n e_{ma} + e_{na} \partial_m \lambda^n) + e_a{}^n \partial_{[m} \lambda_{n]} - e_a{}^{\tilde{n}} \partial_m \lambda^{\tilde{n}}$$
$$\delta e_a{}^m = (\lambda^n \partial_n e_a{}^m - e_a{}^n \partial_n \lambda^m)$$
$$\delta e_a{}^{\tilde{m}} = \lambda^n \partial_n e_a{}^{\tilde{m}} - e_a{}^n \partial_n \lambda^{\tilde{m}}$$

$$\Rightarrow \quad g^{mn} = g^{ab} e_a{}^m e_b{}^n, \quad A_m{}^{\tilde{n}} = e_m{}^a e_a{}^{\tilde{n}}, \quad b_{mn} = \tfrac{1}{2} e_{[m a} e_{n]}{}^a$$

(For the other chirality $\tilde{E}$, $g^{mn}$ has a similar expression, but it is not so easily inverted because of the larger range of the GL indices, so the other expressions are a little more complicated.)

The construction of the action goes as before. The original action [14] now has an additional $F^2$ term for the abelian vectors, and the $b$ field strength is modified because of its altered gauge transformation law:

$$L = \phi^2 (\mathring{R} - \tfrac{1}{12} H^2 - \tfrac{1}{4} g^{mp} g^{nq} F_{mn}{}^{\tilde{r}} F_{pq}{}^{\tilde{r}}) + 4 g^{mn} (\partial_m \phi)(\partial_n \phi)$$
$$H_{mnp} = \tfrac{1}{2} \partial_{[m} b_{np]} + \tfrac{1}{4} A_{[m}{}^{\tilde{q}} F_{np]}{}^{\tilde{q}}, \quad F_{mn}{}^{\tilde{p}} = \partial_{[m} A_{n]}{}^{\tilde{p}}$$

The only identity for the duality invariant objects $f$ and $F$ which differs from that for the ordinary closed string is

$$f_{[ab][cd]} = (c_{ab}{}^e f_{[cd]e} + c_{cd}{}^e f_{[ab]e}) - c_{ab}{}^e c_{cde} - \tfrac{1}{2} F_{ab}{}^{\tilde{m}} F_{cd}{}^{\tilde{m}}$$
$$\Rightarrow \quad F_{[ab][cd]} = -T_{ab}{}^e T_{cde} - \tfrac{1}{2} F_{ab}{}^{\tilde{m}} F_{cd}{}^{\tilde{m}}$$

As a result, the final expression for the manifestly duality and GL invariant action is *the same* as before, since the $F$ term which before contained just the $T^2$ term now contains also the new gauge vector kinetic term, and $H$ as defined in terms of $c$ and $f$ already includes the $AF$ term:

$$S = \int d^D x \left\{ 4 \left[ \nabla \Phi + \tfrac{1}{2} (1 \cdot \overleftarrow{\nabla}) \Phi \right]^2 + \Phi^2 (F^a{}_{[a}{}^b{}_{b]} + F^{ab}{}_{[ab]}) \right\}$$

This form is therefore simpler than the usual form, since duality has automatically included all dependence on the gauge vectors without the addition of any new terms.

## ACKNOWLEDGMENT

I thank Martin Roček for many helpful discussions and suggestions.